\documentclass[letterpaper,aps,prd,preprint,nofootinbib,superscriptaddress]{revtex4}
\pdfoutput=1
\usepackage{graphicx,multirow,subfig}
\usepackage{float}
\usepackage{bm}
\usepackage{braket}
\usepackage{amsmath}
\usepackage{mathtools}
\usepackage{amssymb}
\usepackage{amscd}
\usepackage{latexsym}
\usepackage{slashed}
\usepackage{color}
\usepackage{graphicx}
\usepackage[normalem]{ulem}
\usepackage{color}
\definecolor{orange}{cmyk}{0,0.5,1,0}
\usepackage{verbatim}
\usepackage{rotating}
\usepackage{diagbox}
\usepackage{cancel}
\usepackage[utf8]{inputenc}
\usepackage{array}
\usepackage{caption}

\newcommand{\lagr}{\mathcal{L}}
\newcommand{\URxUBL}[0]{\ensuremath{U(1)_{R} \times U(1)_{B-L}}}

\newcolumntype{L}[1]{>{\raggedright\let\newline\\\arraybackslash\hspace{0pt}}m{#1}}
\newcolumntype{C}[1]{>{\centering\let\newline\\\arraybackslash\hspace{0pt}}m{#1}}
\newcolumntype{R}[1]{>{\raggedleft\let\newline\\\arraybackslash\hspace{0pt}}m{#1}}

\def\be{\begin{equation}}
\def\ee{\end{equation}}
\def\bea{\begin{eqnarray}}
\def\eea{\end{eqnarray}}

\newcommand{\xdownarrow}[1]{%
	{\left\downarrow\vbox to #1{}\right.\kern-\nulldelimiterspace}
}

\begin{document}
\title{${SO(10)}$ inspired $Z'$ models at the LHC}

\author{Simon J.D. King}
\email[]{sjd.king@soton.ac.uk}
\affiliation{\small School of Physics and Astronomy, University of Southampton,
	Southampton, SO17 1BJ, United Kingdom}
\author{Stephen F. King}
\email[]{king@soton.ac.uk}
\affiliation{\small School of Physics and Astronomy, University of Southampton,
	Southampton, SO17 1BJ, United Kingdom}
\author{Stefano Moretti}
\email[]{s.moretti@soton.ac.uk}
\affiliation{\small School of Physics and Astronomy, University of Southampton,
	Southampton, SO17 1BJ, United Kingdom}

\date{\today}
\begin{abstract}
	\footnotesize
We study and compare various $Z'$ models arising from $SO(10)$, focussing in particular on the 
Abelian subgroup $U(1)_{R} \times U(1)_{B-L}$,
broken at the TeV scale to Standard Model hypercharge $U(1)_{Y}$.
The gauge group $U(1)_{R} \times U(1)_{B-L}$, which is equivalent to the $U(1)_{Y}\times U(1)_{\chi}$ in a different basis,
is well motivated from $SO(10)$ breaking and allows neutrino mass via the linear seesaw mechanism.
Assuming supersymmetry, we consider single step gauge unification to predict the gauge couplings, then consider the detection and characterisation prospects of the resulting $Z'$ at the LHC by studying its possible decay modes into di-leptons as well as into Higgs bosons.
The main new result here 
is to analyse in detail the expected leptonic forward-backward asymmetry 
at the high luminosity LHC and show that it may be used to discriminate the $U(1)_{R} \times U(1)_{B-L}$ model
from the usual $B-L$ model based on $U(1)_{Y}\times U(1)_{B-L}$.
		\end{abstract}
\maketitle

\section{Introduction}
$SO(10)$ Grand Unified Theories (GUTs) are very attractive since they predict right-handed neutrinos and
	make neutrino mass inevitable. Supersymmetry (SUSY) allows for a single step unification of the gauge couplings.
	Being a rank 5 gauge group, $SO(10)$ also naturally accommodates an additional $Z'$ gauge boson, which may have a mass at the TeV scale within the range of the Large Hadron Collider (LHC). 
	Such $Z'$ models are attractive since, apart from the three right-handed neutrinos, 
	they do not require any new exotic particles to make the theory anomaly free.

There are two main symmetry breaking patterns of $SO(10)$ leading to the Standard Model (SM) gauge group.
Firstly there is the $SU(5)$ embedding,
	\begin{equation}
SO(10) \rightarrow SU(5)\times U(1)_{\chi} \rightarrow  SU(3)_C\times SU(2)_L \times U(1)_{Y}\times U(1)_{\chi}
\label{su5}
\end{equation}
where the $U(1)_{\chi}$ is broken at the TeV scale,
yielding a massive $Z'_{\chi}$. For recent examples of models based on such a $Z'_{\chi}$, see e.g. \cite{King:2017anf}.

Secondly there is the Pati-Salam
	gauge group embedding,
	\begin{equation}
SO(10) \rightarrow SU(4)_{PS}\times SU(2)_L \times SU(2)_{R}  
\end{equation}
		The Pati-Salam colour group $SU(4)_{PS}$ may be broken to $SU(3)_C\times U(1)_{B-L}$,
	leading to the left-right symmetric model gauge group.
The $SU(2)_{R}$ group may be broken to the gauge group $U(1)_R$ associated with the diagonal generator $T_{3R}$.
It is thus possible to break $SO(10)$ in a single step at the GUT scale without reducing the rank,
\begin{equation}
SO(10) \rightarrow  SU(3)_C\times SU(2)_L \times U(1)_{R}\times U(1)_{B-L}
\label{BLR00}
\end{equation}
The resulting gauge group in Eq.\ref{BLR00} does not predict any new charged currents and 
is not very tightly constrained phenomenologically.
It may therefore survive down to the TeV scale before being broken to the SM gauge group,
leading to the prediction of a massive $Z'_{BLR}$, accessible to the Large Hadron Collider (LHC).

In this paper we shall focus on $SO(10)$ broken at the GUT scale in a single step,
as in Eq.\ref{BLR00}. In order to allow for gauge coupling unification we shall assume supersymmetry (SUSY) 
which is broken close to the TeV scale, but at a high enough scale
to enable the superpartners to have evaded detection at the LHC.
We shall be interested in the $Z'_{BLR}$ which emerges when the Abelian subgroup $U(1)_{R}\times U(1)_{B-L}$
is broken down to the SM hypercharge gauge group $U(1)_Y$ near the TeV scale (for brevity we refer to this scenario as the BLR model).
We study the discovery prospects of such a $Z'_{BLR}$ at the LHC, its possible decay mode into Higgs bosons, and the expected 
forward-backward asymmetry, comparing the predictions to the well studied $B-L$  model 
based on $U(1)_Y\times U(1)_{B-L}$ \cite{King:2004cx,Khalil:2007dr,FileviezPerez:2010ek,OLeary:2011vlq,Elsayed:2011de}.
We comment on the $U(1)_Y\times U(1)_{\chi}$ model \cite{Arcadi:2017atc,Accomando:2010fz} below.

The Abelian gauge group $U(1)_{R}\times U(1)_{B-L}$ has quite a long history in the literature
as reviewed in \cite{Langacker:2008yv,Accomando:2010fz}. 
It was recently realised that SUSY $SO(10)$ models which break down to this gauge group
may allow for a new type of seesaw model, namely the linear seesaw model \cite{Malinsky:2005bi,Hirsch:2009mx}.
Subsequently, the phenomenology of the SUSY $U(1)_{R}\times U(1)_{B-L}$ model has been studied 
in a number of works \cite{DeRomeri:2011ie,Hirsch:2011hg,Hirsch:2012kv,Basso:2012ew,Frank:2017ohg,Chakrabortty:2009xm,Krauss:2013jva}.
Indeed it has been demonstrated that the Abelian BLR gauge group $U(1)_{R}\times U(1)_{B-L}$
is equivalent to $U(1)_{Y}\times U(1)_{\chi}$ (arising from the breaking chain in Eq.\ref{su5}) 
by a basis transformation and furthermore that this equivalence is preserved under RGE running,
when kinetic mixing is consistently taken into account \cite{Krauss:2013jva}. 
Therefore the physics of the TeV scale $Z'_{BLR}$ considered here should be identical to that of the $Z'_{\chi}$  \cite{Krauss:2013jva}.

We emphasise that there are several new aspects of our study including: 
the statistical significance of producing a $Z'_{BLR}$ at the LHC including finite width and interference effects
(the LHC uses a narrow width approximation);
the study of Higgs final states in the $U(1)_{B-L}\times U(1)_R$ model;
and the study of forward-backward asymmetry at the high luminosity LHC as a discriminator between 
the $U(1)_{R} \times U(1)_{B-L}$ model (or equivalently the $U(1)_{Y}\times U(1)_{\chi}$ model)
and the usual $Z'_{BL}$ based on $U(1)_{Y}\times U(1)_{B-L}$, i.e. the commonly studied $B-L$ model
 \cite{King:2004cx,Khalil:2007dr,FileviezPerez:2010ek,OLeary:2011vlq,Elsayed:2011de}.

The layout of the remainder of the paper is as follows. In section~\ref{model} we discuss the BLR model.
In section~\ref{Zpcouplings} we give the $Z'_{BLR}$ couplings to fermions in this model,
while in section~\ref{ZpHiggs} we give the $Z'_{BLR}$ couplings to Higgs.
In section~\ref{RGE} we present a renormalisation group analysis of the BLR model.
In section~\ref{results} we present the results for the Drell-Yan production of the $Z'$ in the BLR model,
assessing the discovery potential at the LHC, present the leptonic forward-backward asymmetry
as a discriminator of different models, and discuss the Higgs final state branching fractions of $Z'_{BLR}$ decays.
Section~\ref{conclusion} concludes the paper.

\section{Model}
\label{model}

We shall not consider the high energy $SO(10)$ breaking here, so 
the starting point of the considered model is to assume that, 
below the GUT scale, we have the gauge group as on the right-hand side of 
Eq.\ref{BLR00}, namely,
\begin{equation}
SU(3)_C\times SU(2)_L \times U(1)_{R}\times U(1)_{B-L}
\label{BLR0}
\end{equation}
Note that in this basis the hypercharge gauge group $U(1)_Y$ of the SM is not explicitly
present, instead it is ``unified'' into $U(1)_{R}\times U(1)_{B-L}$.
Note that, although the Abelian factors are equivalent to the $U(1)_{Y}\times U(1)_{\chi}$ model by a basis transformation,
we shall work in the $U(1)_{R}\times U(1)_{B-L}$ basis.
In order to allow gauge coupling unification we need SUSY, but we shall assume it is broken above the $Z'_{BLR}$ mass scale
so that SUSY particles are not present in the decays of the $Z'_{BLR}$.
Note that such SUSY decays have been considered extensively in 
\cite{DeRomeri:2011ie,Hirsch:2011hg,Hirsch:2012kv,Basso:2012ew,Frank:2017ohg,Chakrabortty:2009xm,Krauss:2013jva}.

At the $Z'_{BLR}$ mass scale (typically a few TeV), hypercharge emerges from the breaking,
\begin{equation}
U(1)_{R}\times U(1)_{B-L} \rightarrow U(1)_Y
\label{BLR}
\end{equation}
where the hypercharge generator is identified as
\begin{equation}
Y=T_{3R}+T_{B-L},
\label{Y}
\end{equation}
where 
\begin{equation}
T_{B-L} = (B-L)/2.
\label{TBL}
\end{equation}
The symmetry breaking in Eq.\ref{BLR} requires two Higgs superfields $\chi_{1,2}$ whose scalar components develop Vacuum Expectation
Values (VEVs) which carry non-zero $T_{3R}$ and opposite $T_{B-L}$ so that they  
are neutral under hypercharge. If they arise from an $SU(2)_R$ doublet then this fixes their charges to be $T_{3R}=\pm 1/2$
and hence $T_{B-L}=\mp 1/2$. Two of them with opposite quantum numbers are required by SUSY to cancel anomalies
(and for holomorphicity). 
They must be singlets under both $SU(3)_C$ and $SU(2)_L$ in order to 
preserve these gauge groups.

Finally, at the electroweak (EW) scale we have the usual Standard Model (SM) breaking
\begin{equation}
SU(2)_L \times U(1)_Y \rightarrow U(1)_Q,
\end{equation}
where the electric charge generator is identified as
\begin{equation}
Q=T_{3L}+Y.
\label{Q}
\end{equation}
As in usual SUSY models, the EW symmetry breaking is accomplished by two Higgs doublets $H_{u,d}$
of $SU(2)_L$ which have $B-L=0$. If the two Higgs doublets of $SU(2)_L$ were embedded into a single $SU(2)_R$ doublet,
then we expect that $H_{u,d}$ will have $T_{3R}=\pm 1/2$, respectively.
In addition, in order to accomplish neutrino masses via the linear seesaw model, we need to add three
complete singlet superfields $S$, as discussed in the appendix~\ref{A}.
The particle content of the model (henceforth denoted as BLR)
is then summarised in Tab.~\ref{particlecontent}. 

\begin{table}
	\centering
\begin{tabular}{c | c | c | c | c | c | c }
	Particle & $T_{3L}$ & $T_{3R}$ & $T_{B-L}$ &$T_{\chi}$ & $Y = T_{3R} + T_{B-L}$ & $Q=T_{3L} + Y$\\ \hline
	&&&&&\\[-1em]
	\multirow{2}{*}{
		$\begin{pmatrix}
		u \\ d
		\end{pmatrix}_L$} & +1/2 & 0 & +1/6 &+1/4& +1/6 &+2/3 \\
	& -1/2 & 0 & +1/6 &+1/4& +1/6 &-1/3 \\
	
	$u_R$ & 0 & +1/2 & +1/6 &-1/4& +2/3 &+2/3 \\
	
	$d_R$ & 0 & -1/2 & +1/6 &+3/4& -1/3 &-1/3 \\
	
	&&&&&\\[-1em]
	\multirow{2}{*}{
		$\begin{pmatrix}
		\nu _e \\ e^-
		\end{pmatrix}_L$} & +1/2 & 0 & -1/2 &-3/4& -1/2 &0 \\
	& -1/2 & 0 & -1/2 &-3/4& -1/2 &-1\\
	
	$\nu_R$ & 0 & +1/2 & -1/2 &-5/4& 0 &0 \\

	$e_R$ & 0 & -1/2 & -1/2 &-1/4& -1 &-1 \\
	
	$\chi _R ^1$&0&-1/2&+1/2&+5/4&0&0\\
	
	$\chi _R ^2$&0&+1/2&-1/2&-5/4&0&0\\
	
	$S$&0&0&0&0&0&0\\
	
	&&&&&&\\[-1em]
\multirow{3}{*}{$H\begin{cases}
		H_u =\begin{pmatrix}
			\phi_u ^+ \\ \phi_u ^0
		\end{pmatrix}_L\\[-1em]
		\\
		H_d= \begin{pmatrix}
		\phi_d ^0 \\ \phi _d ^-
		\end{pmatrix}_L		
	\end{cases}$}&+1/2&+1/2&0&-1/2&+1/2&+1\\[-1.2em]
&&&&&&\\
	&-1/2&+1/2&0&-1/2&+1/2&0 \\
	&&&&&&\\[-0.2em]
	&+1/2&-1/2&0&+1/2&-1/2&0\\
	&&&&&&\\[-1.2em]
	&-1/2&-1/2&0&+1/2&-1/2&-1\\
		\end{tabular}
\caption{\footnotesize The particle content and generators of the $SU(3)_C\times SU(2)_L \times U(1)_{R}\times U(1)_{B-L}$ model.}
	\label{particlecontent}
\end{table}

\section{$Z'$ couplings to fermions}
\label{Zpcouplings}
In this work, numerically, we use the SARAH program \cite{Staub:2013tta} to determine the vector and axial couplings of the fermions with the $Z'_{BLR}$. This includes the full impact of Gauge-Kinetic Mixing (GKM) as done in \cite{Hirsch:2012kv,Krauss:2013jva}. Considering this effect in full leads to $\sim \mathcal{O}(1) \%$ differences in vector and axial couplings. In this section, for simplicity, we neglect the impact of GKM but stress that all implications are considered in our final results.

We begin by examining the low energy breaking of the gauge group in Eq.\ref{BLR}.
The coupling of a fermion $f$ to the $U(1)_R$ and $U(1)_{B-L}$ fields are obtained from 
\begin{equation}
-\lagr _{\rm BLR} = \bar{f} \gamma^\mu \left( g_R T_{3R} W^3 _{\mu R} + g_{BL} T_{B-L} B^{BL} _\mu \right) f,
\end{equation}
where $T_{B-L}=\frac{B-L}{2}$. 

After symmetry breaking, these two fields will mix to become the SM massless hypercharge gauge boson, $B_\mu$, and a massive 
$Z'_\mu$ (corresponding to the $Z'_{BLR}$),
	\begin{equation}	
	\begin{pmatrix}
	B^{BL} _\mu \\
	W^3 _{\mu R} \\
	\end{pmatrix}
	=
	\begin{pmatrix}
	\cos \theta_{BL} & -\sin \theta_{BL} \\
	\sin \theta_{BL} & \cos \theta_{BL} \\
	\end{pmatrix}
	\begin{pmatrix}
	B_\mu \\
	Z^\prime _\mu
	\end{pmatrix}.
	\end{equation}
So, the $Z'_{BLR}$ has the following coupling to fermions:
\begin{equation}
-\lagr ^{Z'}_{\rm BLR} = Z' _\mu \bar{f} \gamma ^\mu \left( g_R \cos \theta _{BL}  T_{3R} - g_{BL} \sin \theta _{BL} T_{B-L} \right) f.
\end{equation}
Since
\begin{equation}
g_{R}\sin \theta _{BL} = g_{BL} \cos \theta _{BL} = g_Y ,
\end{equation}
we may rewrite the $Z'$ couplings of the BLR model in a more compact form,
\begin{align}
& -\lagr ^{Z'}_{\rm BLR} =  Z' _\mu \bar{f} \gamma ^\mu g_Y Q_{LR} f, \nonumber \\
& Q_{LR} \equiv  \left(\cot  \theta _{BL} T_{3R} - \tan  \theta _{BL} T_{B-L} \right), \ \ \ \tan  \theta _{BL} = g_{BL}/g_R.
\label{ZpBLR}
\end{align}

We shall be interested in comparing the $Z'$ couplings in the BLR model above to those in related models
where the SM gauge group (including hypercharge) is augmented by an Abelian gauge group $U(1)'$, identified with the generator $T_{BL}$, resulting in the $Z'$ couplings
\begin{equation}
-\lagr ^{Z'}_{BL} =Z' _\mu \bar{f} \gamma ^\mu g_{BL}T_{B-L} f,
\label{ZpBL}
\end{equation}
which may be compared to the BLR couplings in Eq.\ref{ZpBLR}.
%In fact, as noted in \cite{Langacker:2008yv}, since the generator $T_{\chi}$ is orthogonal to hypercharge $Y=T_{3R}+T_{B-L}$, it 
%may be expressed as
%\begin{equation}
%T_{\chi} =\sqrt{\frac{3}{5}}  \left(\sqrt{\frac{2}{3}}  T_{3R} - \sqrt{\frac{3}{2}} T_{B-L} \right).
%\label{Tchi}
%\end{equation}
%In other words, $T_{\chi}$ corresponds to the generator $\sqrt{3/5}Q_{LR}$ in Eq.\ref{ZpBLR} for $\tan  \theta _{BL}=\sqrt{3/2}$.
%The factor of $\sqrt{3/5}$ is just a GUT normalisation factor. 
%Indeed, at the GUT scale, $T_{\chi}$ and the GUT normalised $Q_{LR}$ are exactly the same since the GUT normalised
%couplings are unified and $\tan  \theta _{BL}=\sqrt{3/2}$ is predicted, as discussed later.
%However the low energy couplings $g_R$ and $g_{B-L}$ are determined by the renormalisation group equations
%(RGEs), as we discuss
%in a later section. 
We shall find to one-loop the non-GUT normalised couplings (i.e., in the conventions of this section)\footnote{Including GUT normalisation, $\sqrt{3/2}g_{BL}=0.563$. We also find the mixed couplings, related to GKM, $g_{R,BL}\sim g_{BL,R} \sim 0.01$.}:
\begin{equation}
g_R=0.448, \ \ 
g_{BL}=0.459. \ \ 
%\tan  \theta _{BL} =1.33, \ \ 
%\cot  \theta _{BL} =0.755.
\label{couplings}
\end{equation}
%This implies that there will be a small difference between the BLR couplings and the $\chi$ couplings due to RGE running
%leading to a slight deviation from $\tan  \theta _{BL}=\sqrt{3/2}=1.33$.

In general the $Z'_{BLR}$ couples to a fermion $f$ which may be either left- or right-handed and the above couplings sum over both
chiral components of all the fermions. For analysing the couplings of different models it is useful to decompose the couplings
into either left-chiral or right-chiral components, leading to the vector and axial couplings in the BLR model as follows
\begin{equation}
-\lagr ^{Z'}_{\rm BLR} =
g_YZ' _\mu \bar{f} \gamma ^\mu (\epsilon _L ^f P_L + \epsilon _R ^f P_R )f = g_YZ' _\mu \bar{f} \gamma ^\mu \frac{1}{2}\left( {g_V ^f - g_A ^f \gamma ^5 }\right)  f,
\label{epsgVA}
\end{equation}
where $P_{R,L} = (1 \pm \gamma _5 ) /2$ and the vector/axial couplings are defined as $g_{V/A} ^f = \epsilon _L ^f \pm e_R ^f$. 
Similar decompositions can be made for the $Z'$ couplings of the other models in Eq.\ref{ZpBL}.
Tab.  \ref{tab:epsilon} shows the chiral couplings for the relevant generators $T_R$ and $T_{B-L}= (B-L)/2$.
Tab. \ref{tab:vector_axial} shows the vector and axial couplings obtained for the two different models.

\begin{table}
	\centering
	\begin{tabular}{c | c | c | c | c | c | c | c | c}
		Model &$\epsilon_L ^u$ & $\epsilon_R ^u$ & $\epsilon_L ^d$ & $\epsilon_R ^d$ & $\epsilon_L ^e$ & $\epsilon_R ^e$ & $\epsilon_L ^\nu$ & $\epsilon_R ^\nu$ \\ \hline
		
		$T_{3R}$ & $0$ & $1/2$ & $0$ & $-1/2$ & $0$ & $-1/2$ & $0$ & $1/2$ \\
		
		$T_{B-L}$ & $1 /6$ & $1 /6$ & $1 /6$ & $1 /6$ & $-1/2$ & $-1/2$ & $-1 /2$ & $-1 /2$ 
%			\\	
%		$T_{\chi}$ & $-1/\sqrt{40}$ & $1/\sqrt{40}$ & $-1/\sqrt{40}$ & $-3/\sqrt{40}$ & $3/\sqrt{40}$ & $1/\sqrt{40}$ & $3/\sqrt{40}$ & $5/\sqrt{40}$
	\end{tabular}
	\caption{\footnotesize Chiral couplings for the $U(1)_R$ and $U(1)_{B-L}$ models.}
	\label{tab:epsilon}
\end{table}
\begin{table}
	\centering
	\begin{tabular}{c | c | c | c | c | c | c | c | c }
		
		Model  &  $g_V ^u$ & $g_A ^u$ & $g_V ^d$ & $g_A ^d$ & $g_V ^e$ & $g_A ^e$ & $g_V ^\nu$ & $g_A ^\nu$ \\ \hline
		
		$T_{3R}$ & $1 /2$ & $-1/2$ & $-1 /2$ & $1/2$ & $-1/2$ & $1/2$ & $0$ & $0$ \\
		
		$T_{B-L}$  & $1 /3$ & $0$ & $1 /3$ & $0$ & $-1$ & $0$ & $-1 /2$ & $-1/2$ 
%		\\
%				
%		$T_{\chi}$ &0&$-2/\sqrt{40}$&$-4/\sqrt{40}$&$2/\sqrt{40}$&$4/\sqrt{40}$&$2/\sqrt{40}$&$3/\sqrt{40}$&$3/\sqrt{40}$\\
		\end{tabular}
	\caption{\footnotesize Vector and axial couplings for the $U(1)_R$ and $U(1)_{B-L}$ models. Note that we have integrated out the right-handed neutrinos$^2$ in calculating $g_V ^\nu$ and $g_A ^\nu$.}
	\label{tab:vector_axial}
\end{table}
\addtocounter{footnote}{1}
\footnotetext{In the linear seesaw, the heavy neutrino mass is approx $M_{N} \sim \tilde{F} v_R$, see Eq.\ref{eq:App1} in appendix A for the definition of $\tilde F$ while $v_R$ is the
BLR breaking scale. We will see that the mass of the $Z'$ is approximately
	$M_{Z'} \sim \frac{1}{2} \sqrt{\left(\frac{3}{2} g_{B-L}^2 + g_R ^2 \right)}v_R $. We thus prevent  heavy neutrino decays $(2M_N > M_{Z'})$ through the requirement that the free Yukawa coupling be large enough, $\tilde{F} > \sqrt{\left(\frac{3}{2} g_{B-L}^2 + g_R ^2 \right)} \sim 0.2$.}

\section{$Z'$ Couplings to Higgs Bosons }
\label{ZpHiggs}
In this section we shall ignore the $Z'_{BLR}$ decays into bosons arising from $\chi_R^1$ and $\chi_R^2$.
The $\chi_R^1$ and $\chi_R^2$ bosonic sector contains four degrees of freedom, two scalars plus two pseudoscalars,
where one of the pseudoscalars is eaten by the $Z'_{BLR}$, to leave two $CP$ even scalars plus one $CP$ odd pseudoscalar in the physical
spectrum. If the soft SUSY breaking masses associated with $\chi_R^1$ and $\chi_R^2$ are very large, 
then we would expect the 
physical $CP$ odd pseudoscalar to become very heavy.
Since the $Z'_{BLR}$ must decay into a scalar plus a pseudoscalar (assuming that $CP$ and angular momentum are conserved) then this would imply that 
none of the bosons arising from $\chi_R^1$ and $\chi_R^2$ would be kinematically accessible in $Z'_{BLR}$ decays.

Under the above assumption of large soft masses for $\chi_R^1$ and $\chi_R^2$,
we shall discuss the $Z'_{BLR}$ coupling to the Higgs bosons arising from $H_u$ and $H_d$ only,
which are assumed to have smaller soft masses.
To investigate the $Z'$ coupling to what is essentially a 2-Higgs Doublet Model (2HDM) sector, we begin with the Lagrangian term with the covariant derivative
\begin{equation}
\lagr_{Z',\textrm{scalars}} = (D_\mu \Phi _1) ^\dagger ( D_\mu \Phi _1) + (D_\mu \tilde{\Phi}_2 ) ^\dagger (D_\mu \tilde{\Phi}_2)
\end{equation}
with
\begin{equation}
D_\mu = \partial _\mu - i \frac{g_Y}{s_{BL} c_{BL}} (T_{3R} - s_{BL}^2\frac{Y}{2}),
\end{equation}
where $\cos \left(\theta _{B-L}\right)\equiv c_{BL}$ and $\sin \left(\theta _{B-L}\right)\equiv s_{BL}$. Our two Higgs doublets are
\begin{equation}
\Phi _1 = \begin{pmatrix}
\phi_1 ^+ \\
(v_1 + h_1 + ia_1)/\sqrt{2}
\end{pmatrix},
~~
\tilde{\Phi}_2=i \sigma_2 \Phi _2 ^* =
\begin{pmatrix}
\phi_2 ^+ \\
(-v_2 - h_2 + ia_2)/\sqrt{2}
\end{pmatrix}
\end{equation}
and we rotate the fields to the physical basis as in the standard 2HDM procedure,
\begin{equation}
\Phi _1 ^R = \begin{pmatrix}
G ^+ \\
(h^0 s_{\beta \alpha} + H^0 c_{\beta \alpha} + v_{SM} + iG^0)/\sqrt{2}
\end{pmatrix},
~~
\tilde{\Phi}_2 ^R =
\begin{pmatrix}
H ^+ \\
(-h^0 c_{\beta \alpha} + H^0 s_{\beta \alpha} + i A^0)/\sqrt{2}
\end{pmatrix},
\end{equation}
where we defined the standard 2HDM rotation angles $\cos (\alpha -\beta ) \equiv c_{\alpha \beta}$ and $\sin (\alpha -\beta ) \equiv s_{\alpha \beta}$. We extract the physical couplings for our $Z'_{BLR}$ to the $h^0, H^0, H^\pm ,A^0$ in
Tab. \ref{tab:Zp2HDM}.
\begin{table}[h]
	\centering
\begin{tabular}{c | c}
	Vertex & $g_{Z' S_1 S_2}$  \\
	\hline
	&\\[-1.5em]
	$Z'h^0 A^0$ & $\dfrac{g_R \cos \theta_{B-L} \cos (\beta - \alpha)}{2} $\\
	&\\[-1.5em]
	$Z'H^0 A^0$ & $\dfrac{- g_R \cos \theta_{B-L} \sin (\beta - \alpha)}{2} $\\
	&\\[-1.5em]
	$Z' H^+ H^-$ &  $-i \dfrac{g_R \cos \theta_{B-L} }{2}$\\
\end{tabular}
\caption{\footnotesize The coupling of the BLR $Z'$ to the physical 2HDM mass states. The Feynman rule for the vertex
is given by  $(g_{Z' S_1 S_2})(p - p') _\mu$, where $p,p'$ are the momenta of the two scalars towards the vertex.}
	\label{tab:Zp2HDM}
\end{table}

We find the partial widths by using the general expression for a $Z'$  decaying into two spinless bosons of unequal masses $M_1$ and $M_2$, with coupling $g_{Z' S_1 S_2}$ (read off from Tab.~\ref{tab:Zp2HDM}),
\begin{equation}
\Gamma (Z'_{BLR} \rightarrow S_1 S_2) = \frac{1}{48 \pi} \frac{1}{M_{Z'}^3}
g^2_{Z' S_1 S_2}
\left(
M_{Z'}^4 +M_1^4+M_2^4 -2 \left(M_2^2 M_{Z'}^2+M_1^2 M_{Z'}^2+M_1^2 M_2^2\right)
\right).
\end{equation}
For a discussion of the $Z'_{BL}$ coupling to the scalar sector in the $U(1)_{B-L}$ model see e.g. \cite{Basso:2012ew}.

\section{Renormalisation Group Equations}
\label{RGE}
We now turn to the Renormalisation Group Equations
(RGEs) at one-loop. These RGEs will determine the $U(1)_R$ and $U(1)_{B-L}$ coupling constants and will 
 also predict a value of the SM hypercharge coupling constant, given measured results of $\alpha_2$ and $\alpha _3$. We begin by using the SM $\beta$-function coefficients $b_2 ^{\rm SM} = -19/6$ and $b_3 ^{\rm SM} =-7$ for the $SU(2)_L$ and $SU(3)_c$ groups, respectively. We perform the running  from $M_Z$ up to our BLR breaking scale, which we denoted by $v_R$. From the scale $v_R < Q < v_{\rm SUSY}$, these two $\beta$-function coefficients are unchanged, as none of the additional BLR particle content has quantum numbers under these two groups. Then, at $v_{\rm SUSY} < Q < M_{\rm GUT}$, we introduce the SUSY partners and the 
$\beta$-function coefficients are modified to $b_2 ^{\rm SUSY} = +1$ and $b_3 ^{\rm SUSY} =-3$. These are the familiar MSSM $\beta$-function coefficients. The strong and weak coupling constants are run until they intersect, which determines $Q=M_{\rm GUT}$ and $\alpha_{\rm GUT} \equiv \alpha _{2} (M_{\rm GUT}) = \alpha _{3} (M_{\rm GUT})$. We now run our $U(1)_{B-L}$ and $U(1)_R$ coupling constants down from this GUT scale. 

\begin{figure}
	\includegraphics[width=0.9\linewidth]{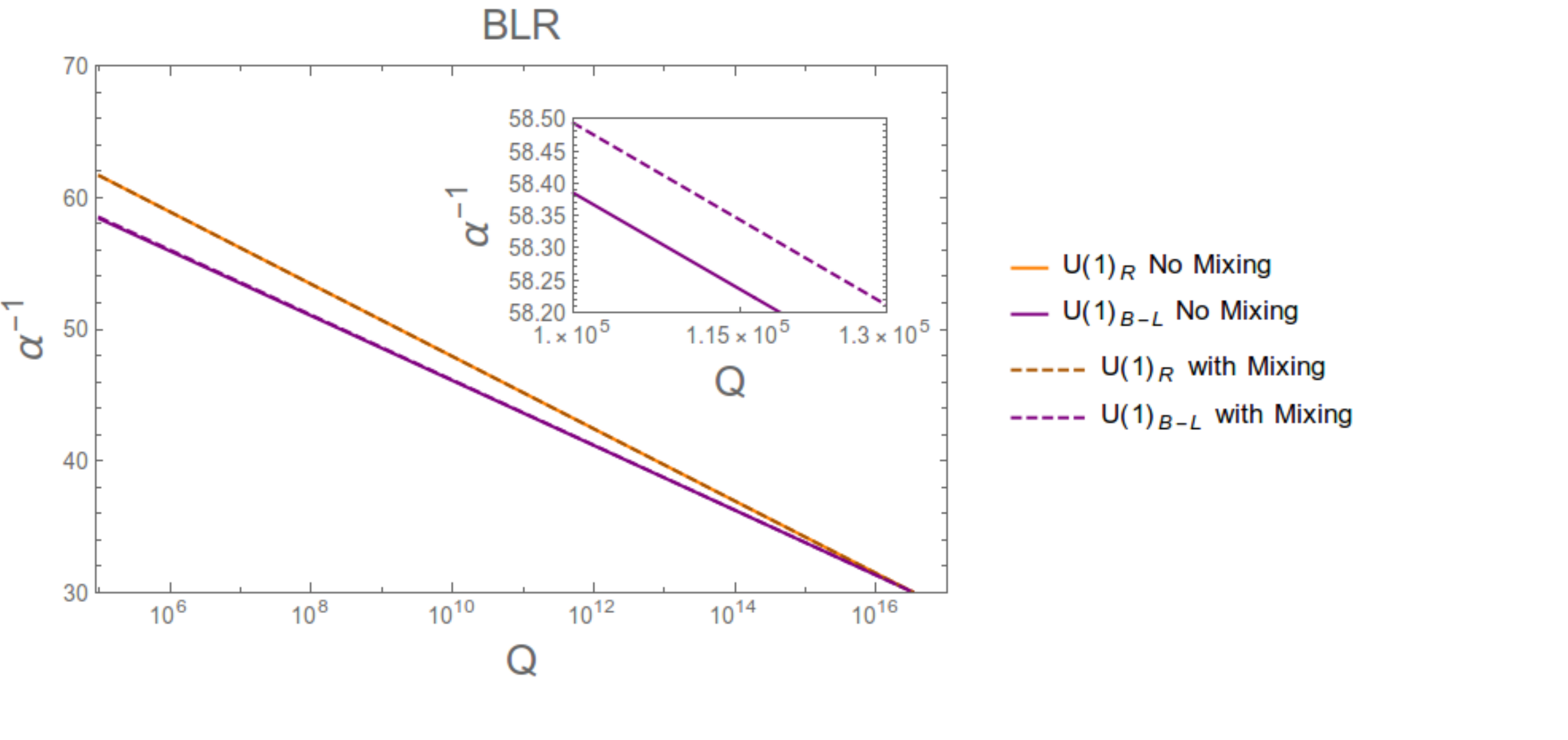}
	\caption{Comparison of RGE evolution with (solid lines) and without (dashed lines) gauge-kinetic mixing from GUT to SUSY scale. The $U(1)_{R}$ evolution is unchanged, whereas the $U(1)_{B-L}$ is modified slightly. A zoomed in plot of this modification is shown.}
	\label{fig:RG_Mix}
\end{figure}

As we have two $U(1)$ groups, they undergo GKM. We begin with the $\beta$-function
coefficients $b_{BL} ^{\rm BLR,SUSY}=27/4$, $b_R ^{\rm BLR,SUSY}=15/2$ and a mixed term $b_{R,B-L} ^{\rm BLR,SUSY}= -\sqrt{3/8}$, including a GUT normalisation term of $3/8$ on the $U(1)_{B-L}$ and hence $\sqrt{3/8}$ on the $( U(1)_{B-L} \times U(1)_R )$ coefficient. Rotating the couplings into the upper triangular physical basis \cite{Coriano:2015sea}, and following the procedure of  \cite{Bertolini:2009qj}, we find the following $\beta$-functions for the GUT normalised couplings\footnote{The couplings in this section are GUT normalised, while those in earlier sections are the non-GUT normalised couplings
We have chosen the same nomenclature for both normalisations, being careful to specify which normalisation we are using.}
\begin{align}
\frac{d g_R}{dt}&=\frac{1}{(4 \pi)^2} \frac{15 g_{R}^3}{2},\\
\frac{d \tilde{g}}{dt} &=\frac{1}{(4 \pi)^2} \left[\left(\frac{27 }{4}g_{BL}^2 -\sqrt{\frac{3}{2}} g_{BL} \tilde{g}+\frac{15}{2}\tilde{g}^2\right)\tilde{g} +\left(-\sqrt{\frac{3}{2}}  g_{BL}+15  \tilde{g}\right)g_{R}^2 \right],\\
\frac{d g_{BL}}{dt}&=\frac{1}{(4 \pi)^2}\left(\frac{27 }{4}g_{BL}^2 -\sqrt{\frac{3}{2}} g_{BL} \tilde{g}+\frac{15}{2}\tilde{g}^2\right) g_{BL}.
\end{align}
At the GUT scale, we set $\tilde{g}=0$ and allow it to run to non-zero values at low scale. Fig. \ref{fig:RG_Mix} shows the running of the $U(1)_{R}$ and $U(1)_{B-L}$ groups both with (solid line) and without (dashed line) including the GKM procedure. One can see immediately that these two lines lie on top of one another, meaning the effect of the GKM is negligible. The $\alpha_R$ has an entirely negligible change and one can see a zoomed plot of the shift in the $\alpha_{BL}$ coefficient, which changes by $\mathcal{O}(0.1 \%)$. At the low (TeV) scale, one finds a negligible mixing coupling term $\tilde{g} \approx 10^{-2}$, nevertheless we include this correction in our numerical work.

\begin{figure}
	\includegraphics[width=0.9\linewidth]{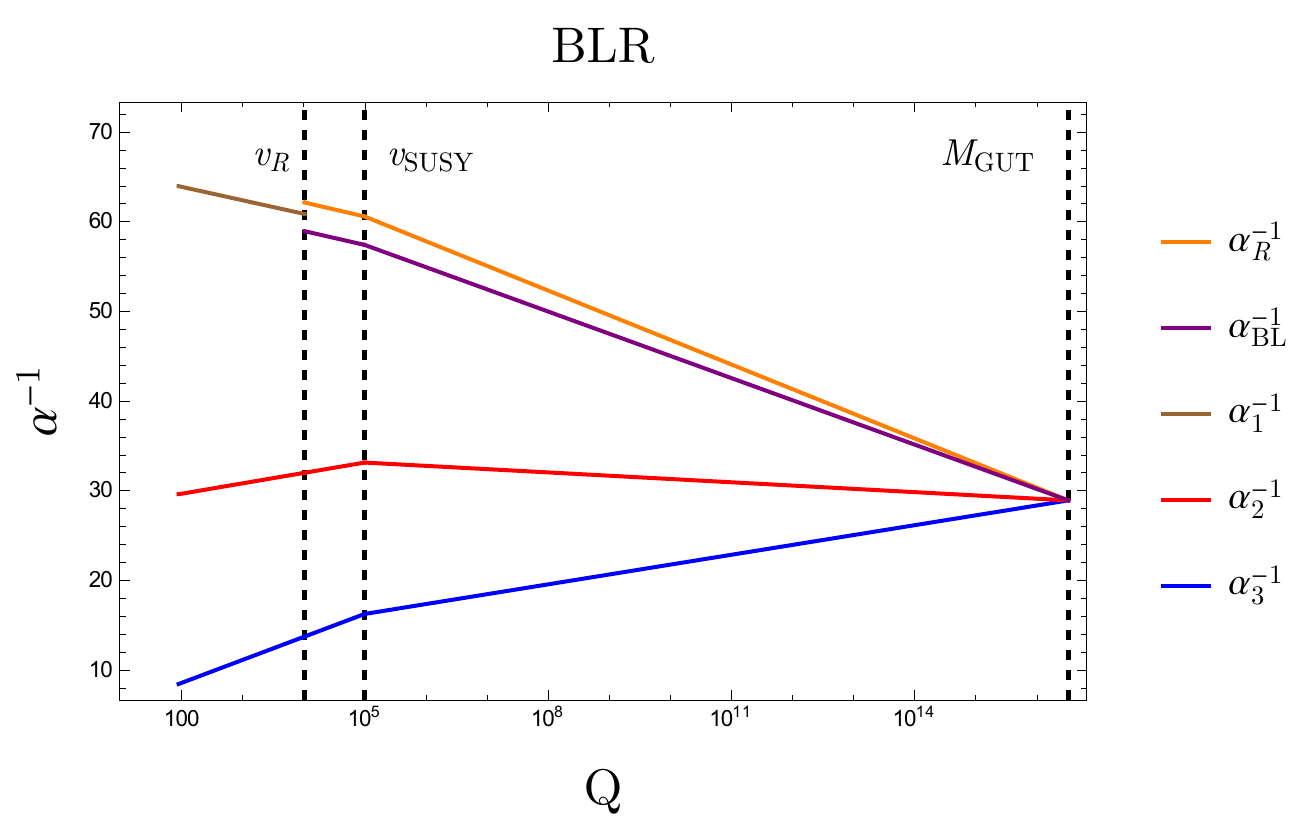}
	\includegraphics[width=0.9\linewidth]{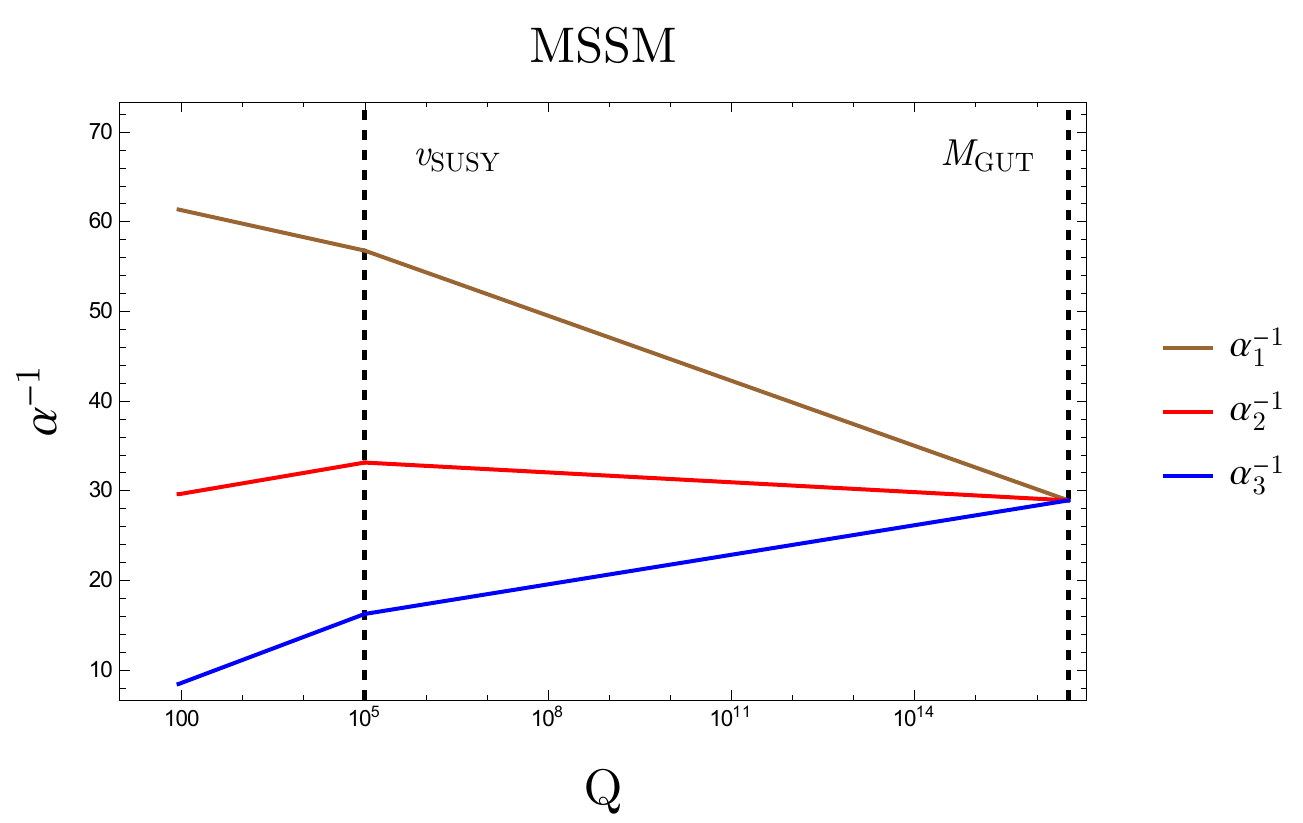}
	\caption{\footnotesize The upper panel shows the running couplings in the BLR model, with $v_R = 11660$ GeV, which corresponds to $M_{Z'}=3750$ GeV and $v_{\rm SUSY}=10^5$ GeV. The GUT scale is determined to be $M_{\rm GUT}=3.30\times 10^{16}$ GeV.
	The lower panel shows the running couplings in the MSSM.}
	\label{fig:RGEGraph}
\end{figure}

%%%
We include GKM from the SUSY scale to the $U(1)_R \times U(1)_{B-L}$ breaking scale, $v_R$. From $v_R < Q < v_{\rm SUSY}$, decoupling the SUSY particles, the $\beta$-function coefficients change to $b_{BL} ^{\rm BLR}=17/4$, $b_R ^{\rm BLR}=13/3$ and a mixed term $b_{R,B-L} ^{\rm BLR,SUSY}= b_{B-L,R} ^{\rm BLR,SUSY}=-1/\sqrt{24}$. We summarise these beta function coefficients and their meaning in appendix \ref{sec:appendix_rge}. At $v_R$ these two coupling values determine the (GUT normalised) hypercharge coupling, \begin{align}
\alpha^{-1} _1 & = \frac{3}{5} \alpha ^{-1} _R + \frac{2}{5} \alpha_{BL}^{-1} .
\end{align}
From this scale, $\alpha_1$ is run further down from $v_R$ to $M_Z$, with the SM $\beta$-function coefficient $b_1 ^{\rm SM}=41/10$. The BLR breaking scale has been chosen such that the VEV  and coupling values at this point correspond to a $Z'$ with a statistical significance $\le 2 \sigma$, which is seen later to be 3750 GeV. Using this $Z'$ mass, the $v_R$ VEV is determined from the formula\footnote{The factor of $3/2$ 
in Eq.\ref{vR} multiplying $g_{B-L}^2$
comes from the $3/8$ GUT normalisation factor times a factor of $4$ in going from $B-L$ to 
$(B-L)/2$. This is responsible for the GUT scale prediction $\tan  \theta _{BL}=g_{BL}/g_R=\sqrt{3/2}$
in terms of the non-GUT normalised couplings in Eq.\ref{ZpBLR}.} \cite{Hirsch:2012kv} in the limit $\tilde{g}=0$,
\begin{equation}
M^2_{Z'}= \frac{1}{4}\left(\frac{3}{2} g_{B-L}^2 + g_R ^2 \right)v_R ^2+\frac{\frac{1}{4}g_R ^4 v^2}{(3/2) g_{B-L}^2 + g_R ^2 }
\approx   \frac{1}{4}\left(\frac{3}{2} g_{B-L}^2 + g_R ^2 \right)v_R ^2   \label{vR},
\end{equation}
where $\sqrt{(3/2)}g_{B-L}=0.563$, as seen in Eq.\ref{couplings}, and $M_{Z'}=3750~\textrm{GeV}$ leads to $v_R = 10328~\textrm{GeV}$. 

The upper panel of 
Fig. \ref{fig:RGEGraph} shows the running couplings of the BLR model,
setting $v_R = 10328$ GeV and $v_{\rm SUSY}=10^5$ GeV. Using our one-loop RGEs, we predict a value for the SM hypercharge coupling as $\alpha_Y (M_Z ) = \dfrac{3}{5} \alpha_1(M_Z) = 1/102.44$, which we may compare to the experimentally determined value of $\alpha_Y ^{\rm exp} = \dfrac{\alpha _{EM}}{1 - \sin ^2 \theta _W}=1/98.39$ \cite{Patrignani:2016xqp}. 
The difference between the two values may be partly accounted for by our procedure of running up the best fit experimental values
of $\alpha_2$ and $\alpha_3$ at $M_Z$ to determine $M_{\rm GUT}$ and $\alpha_{\rm GUT}$ at the point where they meet, 
then running all the gauge couplings from this point down to low energies. This procedure, though convenient for the BLR model
where the hypercharge gauge coupling is not defined above $v_R$, does not take into account the experimental error 
in $\alpha_3^{\rm exp}$ in the prediction for $\alpha_Y ^{\rm exp}$. Another source of error is the fact that we do not consider either two loop
RGEs or threshold effects, both of which are beyond the scope of this paper.
Using our one loop results, we determine the values of the couplings in Eq.\ref{couplings}, which refer to the non-GUT normalised couplings.

For comparison, the lower panel of Fig. \ref{fig:RGEGraph} shows
the MSSM at one-loop running couplings, again assuming $v_{\rm SUSY}=10^5$ GeV.
In this case the analogous procedure to that used in the BLR model yields
a prediction for the SM hypercharge coupling of $\alpha_Y ^{\rm MSSM} (M_Z) = 1/102.25$.

\section{Results}
\label{results}

\subsection{Preliminaries}
In this section, we review the LHC results specific to the BLR model  in Drell-Yan (DY) processes as well as in final states including Higgs bosons. We do so in two separate subsections to follow. In the case of DY studies, we also compare the BLR results to those of the $U(1)_{B-L}$ scenario. Throughout our analysis we assume the aforementioned heavy SUSY scale, thereby preventing decays of the $Z'$ into sparticles.  However, we consider the possibility that the 2HDM-like Higgs states of the BLR models 
are lighter than the $Z'$, which may therefore decay into them via the couplings in Tab.~\ref{tab:Zp2HDM}. Further, notice that $Z'$ decays into non-MSSM-like Higgs states can be heavily suppressed in comparison, in virtue of the fact that the additional CP-odd state not giving mass to the $Z'$ can be made arbitrarily heavy (as previously explained), a setup which we assume here, so that we refrain  from accounting for these decay patterns. Finally,  recall that heavy neutrino decays are prevented here in the light of
footnote 2 and that they have already been studied in, e.g., \cite{Khalil:2015naa} (for the $B-L$ case), from where it is clear that they have little $Z'$ diagnostic power. In contrast, we aim at making the point that the Higgs decays we study below can eventually be used for this purpose. 

We use standard 2HDM notation, 
such that $h^0$ and $H^0$ are the
CP-even Higgs mass states (with the lighter
$h^0$ being the discovered SM-like one), $A^0$ the CP-odd one and $H^\pm$ the charged ones. 

Tab.~\ref{tab:vector_axial2} summarises the numerical values of the vector and axial couplings of  the $Z'$
to fermions for the $B-L$ and BLR models.
For each scenario we have defined new vector and axial couplings with the gauge coupling absorbed:
\begin{equation}
-\lagr^{Z'} = Z_\mu ' \bar{f} \gamma^\mu \frac{1}{2} (\bar{g} ^f _V - \bar{g} ^f _A \gamma^5 )f,
\end{equation}
which may be compared to Eq.~\ref{epsgVA}.
For the $U(1)_{B-L}$ model the calculation of $\bar{g}^f _{V,A}$ in Tab.~\ref{tab:vector_axial2} 
uses the gauge coupling constants shown there 
multiplied by the vector and axial couplings given previously in Tab.~\ref{tab:vector_axial}. For the BLR model, the new numerical vector and axial
couplings are derived including the full effects of gauge-kinetic mixing using SARAH (as a function of the mixed couplings $g_{BL,R} ,~g_{R,BL}$ and the rotation matrix which diagonalises the neutral gauge boson mass matrices), but may be approximated analytically neglecting GKM using Eqs.\ref{ZpBLR},\ref{epsgVA} as
\begin{equation}
\bar{g}^f _{V,A} ({\rm BLR}) \approx g_Y  \left[ (\cot  \theta _{BL}) g^f _{V,A}(R) - (\tan  \theta _{BL}) g^f _{V,A} (BL) \right]
\label{gbar}
\end{equation}
in terms of the vector and axial couplings $g^f _{V,A}(R)$ and $g^f _{V,A} (BL)$ 
for the $T_{3R}$ and $T_{B-L}$ models as written in Tab.~\ref{tab:vector_axial}.
The non-GUT normalised gauge couplings 
for the BLR model in Eq.\ref{gbar} and Tab.~\ref{tab:vector_axial2} come from the RGE analysis
leading to Eq.\ref{couplings}. The values of the non-GUT normalised gauge couplings $g_{BL}$ and $g_{\chi}$ for the $B-L $ and $\chi$ models 
in Tab.~\ref{tab:vector_axial2} were taken from the low energy
parametrisation in \cite{Accomando:2010fz} rather than an RGE analysis, which would require us to specify the corresponding 
high energy models, which we do not wish to do here, bearing in mind that the $B-L$ model does not emerge from $SO(10)$.
If some other value of $g_{BL}$ were used instead, then the vector and axial couplings for the $B-L$ model in Tab.~\ref{tab:vector_axial2} would be straightforwardly rescaled.

Many qualitative features of the results can be understood by examining the fermion couplings
in Tab.~\ref{tab:vector_axial2}, for example, the vector nature of the $B-L$ couplings.

\begin{table}
	\centering
	\begin{tabular}{c | c | c | c | c | c | c | c | c | c}		
Model  &  Gauge Coupling & $\bar{g}_V ^u$ & $\bar{g}_A ^u$ & $\bar{g}_V ^d$ & $\bar{g}_A ^d$ & $\bar{g}_V ^e$ & $\bar{g}_A ^e$ & $\bar{g}_V ^\nu$ & $\bar{g}_A ^\nu$ \\ \hline
\text{$B-L$} &$g_{BL}$=0.592& 0.197 & 0 & 0.197 & 0 & -0.592 & 0 & -0.296 & -0.296 \\
		 \text{BLR} & See Eq.\ref{couplings} & -0.0103 & -0.135 & -0.279 & 0.135 & 0.300 & 0.135 & 0.217 & 0.217 
%		 \\
%		 \text{BLR (No GKM)} & $g_{BL} ^{GN} = 0.459$ & 0 & -0.145 & -0.291 & 0.145 & 0.291 & 0.145 & 0.218 & 0.218
	\end{tabular}
	\caption{\footnotesize Numerical values of the vector and axial couplings for the $U(1)_{B-L}$ and $U(1)_{B-L}\times U(1)_R$ models.
	Note that we have decoupled the right-handed neutrinos in calculating $g_V ^\nu$ and $g_A ^\nu$.}
	\label{tab:vector_axial2}
\end{table}

\begin{figure}
\subfloat[\label{fig:Gauss_Lum_30_Model_BLR}]{\includegraphics[width=0.8\linewidth]{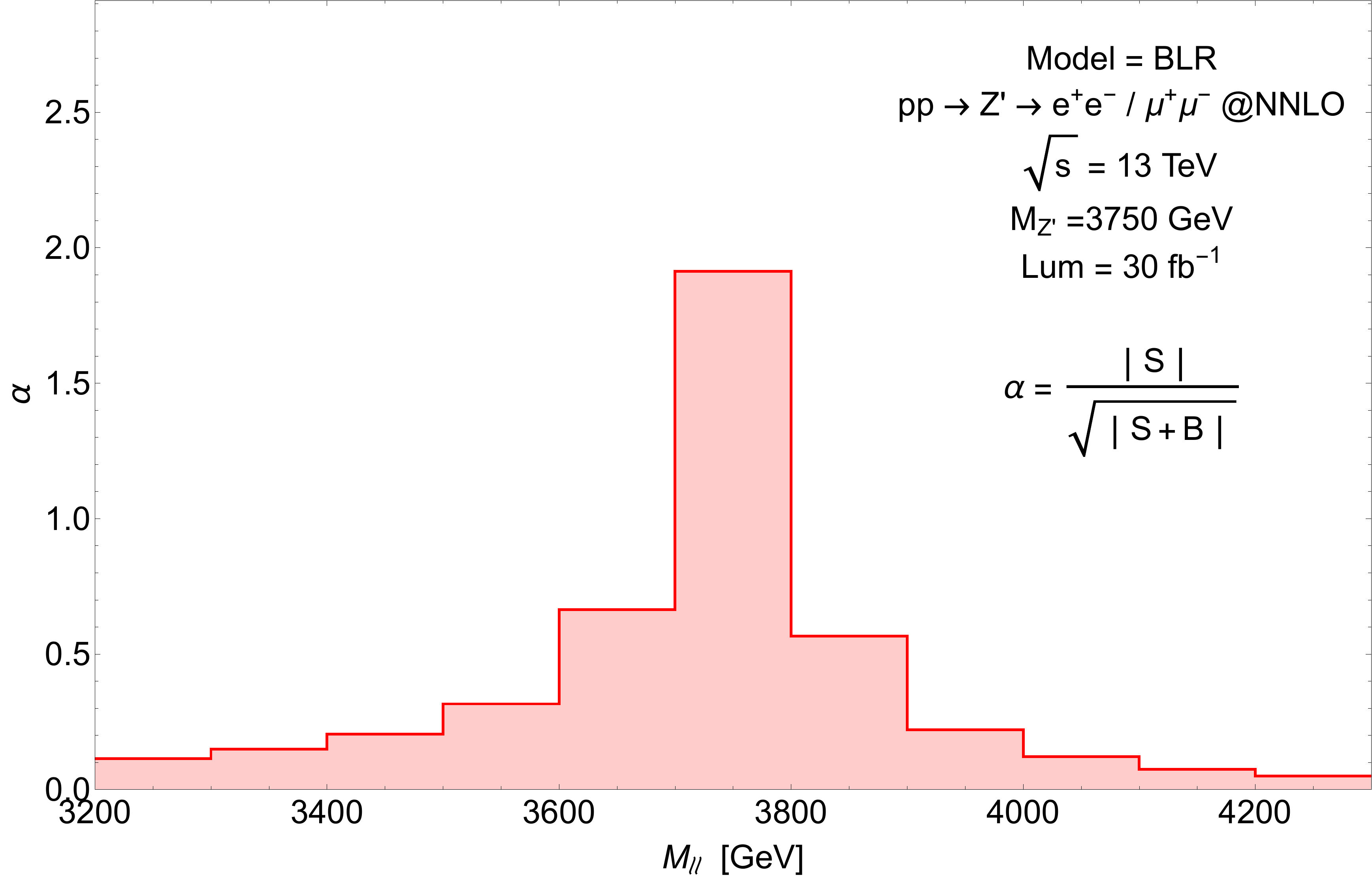}} \\
\subfloat[\label{fig:Gauss_Lum_300_Model_BLR}]{\includegraphics[width=0.8\linewidth]{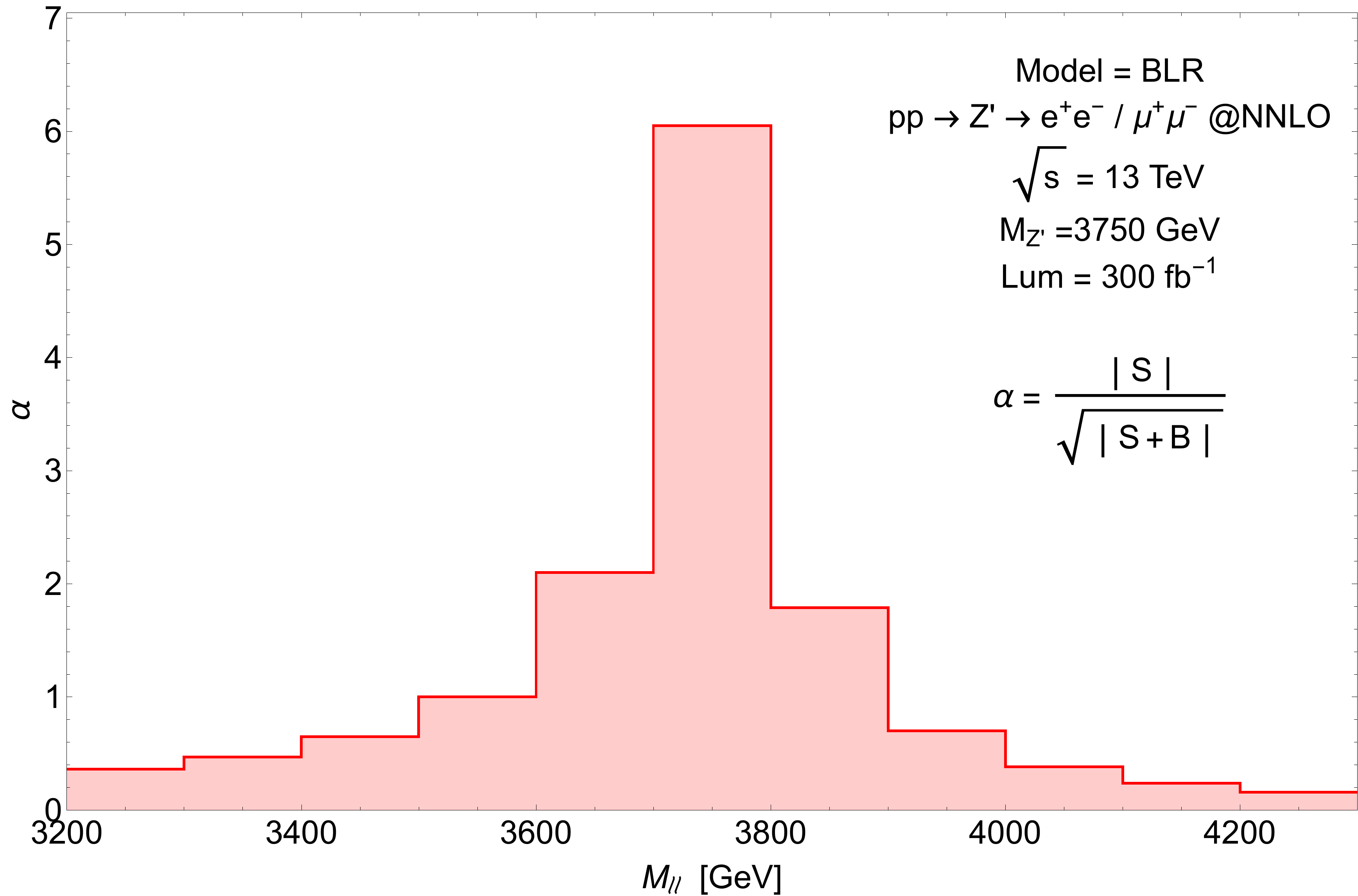}}
\caption{\footnotesize Statistical significance for producing  a $Z'$  decaying into $e^+ e^-$ and $\mu ^+ \mu ^-$ in the BLR model
at integrated luminosities of (a) $L=30$ fb$^{-1}$ and (b) $300$ fb$^{-1}$. The number of events obtained at these luminosities for $pp \to Z'$ is 74 in case (a) and 737 in case (b).}
\end{figure}

\begin{figure}
	\begin{tabular}{c }
		{\includegraphics[width=0.9\linewidth]{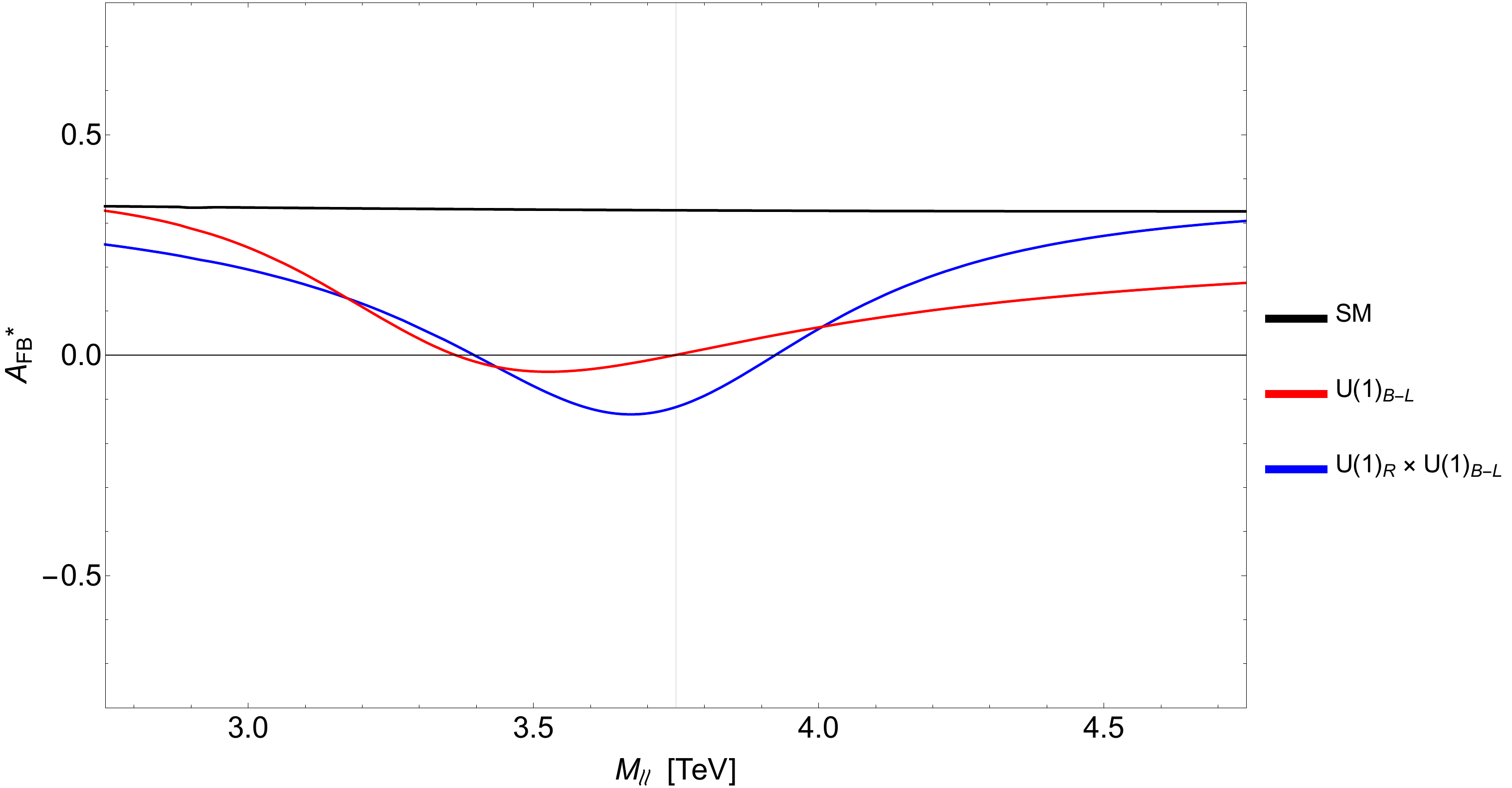}} 
	\end{tabular}
	\caption{\footnotesize The theoretical predictions of the leptonic forward-backward asymmetry at the LHC 
	$A^*_{\rm FB}$ in the presence of a $Z'$  decaying into $e^+ e^-$ and $\mu ^+ \mu ^-$ for the  $U(1)_Y\times U(1)_{B-L}$ (red) and  
	$U(1)_R \times U(1)_{B-L}$ (blue) models. We have taken $M_{Z'}=3750$ GeV. The SM (black) result is also given for comparison.}
\label{fig:AFB}
\end{figure}

%\begin{figure}
%	\begin{tabular}{c }
%		%\subfloat[\label{fig:BR1}]{\includegraphics[width=0.7\linewidth]{BrZp_A0Vary_cosBA_01_Mzp_3750}}\\ 
%		%\subfloat[\label{fig:BR2}]{\includegraphics[width=0.7\linewidth]{BrZp_cosBA_01_Mzp_3750}} 
%	\end{tabular}
%	\caption{\footnotesize BRs of a $Z'$ in the BLR model as a function of degenerate $A_0$, $H^0$ and $H^\pm$ masses. Here, $M_{Z'}=3750$ GeV and $\cos(\beta-\alpha)=0.1$.}
%\end{figure}

\subsection{Drell-Yan}

The most promising channel to search for and profile a $Z'$ boson at the LHC in the BLR model is  DY production and decay, namely, $pp\to \gamma,Z,Z'\to$ $e^+e^-$ and $\mu^+\mu^-$. Fig.~\ref{fig:Gauss_Lum_30_Model_BLR} illustrates the current LHC reach (assuming 30 fb$^{-1}$ of integrated luminosity at 13 TeV), highlighting that a $Z'$ of BLR origin with a mass of 3750 GeV is allowed by data, as its statistical significance $\alpha\equiv\frac{|S|}{\sqrt{|S+B}|}$ is less than 2 in the entire mass range over which the signal $|S|$ could manifest itself over the background $|B|$. Notice that, here and in the following, our signal is given by the  (modulus of the)  cross section of $pp\to \gamma,Z,Z'\to$ $e^+e^-$ and $\mu^+\mu^-$ minus that of  $pp\to \gamma,Z\to$ $e^+e^-$ and $\mu^+\mu^-$  (thereby including   interference effects between $Z'$ and $\gamma,Z$), the latter being the 
(irreducible) background\footnote{ Notice that, for the $Z'$ mass ranges currently allowed by experiment, other (reducible) backgrounds can be neglected.}. This very same $Z'$ boson will, however, become accessible by the end of Run 2 of the LHC, as illustrated in Fig.~\ref{fig:Gauss_Lum_300_Model_BLR}, where (assuming 300 fb$^{-1}$ of integrated luminosity at 13 TeV) values of $\alpha$ in excess of 5 are found near the peak region\footnote{In performing this exercise, we have used the program described in Refs.~\cite{Abdallah:2015hma,Abdallah:2015uba} for the $U(1)_{B-L}$ case suitably adapted to the BLR one. In particular, our implementation accounts for $Z'$ width and interference (with SM di-lepton production) effects, which tend to reduce somewhat the sensitivity of the LHC experiments. Needless to say, when these are neglected, we are able to reproduce results obtained by the LHC collaborations \cite{Aaboud:2016cth,Khachatryan:2016zqb} with percent accuracy, for the corresponding choice of couplings (which differ somewhat from those used in the present paper). This is why our limits
for $Z'$ masses differ from those quoted by the LHC.}.

Once such a $Z'$ signal is established, it will be necessary to diagnose it, i.e., to assess to which model it belongs. A useful variable in this respect is the (reconstructed) Forward-Backward Asymmetry
($A^*_{\rm FB}$) of the DY cross section. We use here the definition adopted in Ref.~\cite{Accomando:2015cfa}, see Sect. 3 therein, with no cut on the the di-lepton rapidity (see also 
Refs.~\cite{Accomando:2015ava,Accomando:2015pqa}).
Fig.~\ref{fig:AFB} shows the shape of this observable at the LHC, for $\sqrt s=13$ TeV and $M_{Z'}=3750$ GeV, as it would appear in the $Z'$ peak region of the di-lepton invariant mass distribution 
for the BLR model as well as the $U(1)_{B-L}$ scenario. The shape emerging from the BLR case is notably different from the one of the companion $SO(10)$ model\footnote{As intimated, recall that the $Z'$ couplings to leptons in the $U(1)_{B-L}$ case are purely vectorial, so that non-zero values of $A^*_{\rm FB}$ are due in this case to interference effects.}.

In order to quantify whether the LHC will be able to differentiate these two models, from one another or the SM, one must include the statistical error in the formulation of $A^*_{\rm FB}$ \cite{Accomando:2015ava}:
\begin{equation}
\delta A_{\rm FB}=\sqrt{\frac{1-A_{\rm FB}^2}{N}}.
\end{equation}

\begin{figure}[h!]
	\includegraphics[width=0.7\linewidth]{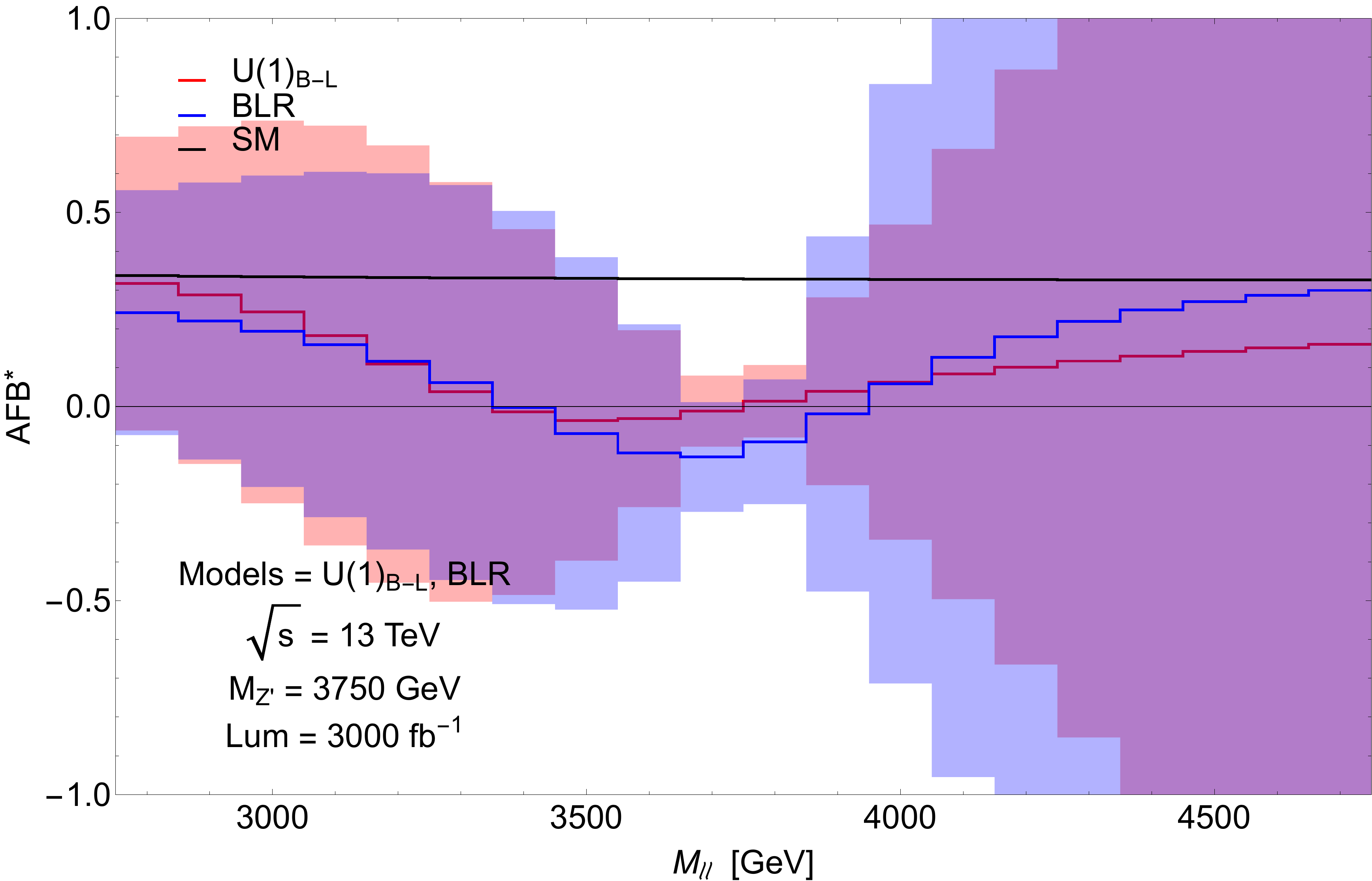}
	\caption{\footnotesize The 
$A^*_{\rm FB}$ spectrum of the DY cross section in the presence of a $Z'$ of mass $M_{Z'}=3750$ GeV.
The figure we shows the BLR model prediction for  $A_{\rm FB}^*$ (in blue) and its error (shaded in light blue) as well as the $U(1)_{B-L}$ prediction for $A_{\rm FB}^*$ (in red) and its error (shaded in light red) as a function of the dilepton invariant mass.
  The purple region is the overlap of errors between the two models. Here, $L=3000$ fb$^{-1}$.}
\label{fig:AFB_U1BL_BLR}
\end{figure}

In Fig.~\ref{fig:AFB_U1BL_BLR} we include this error in a binned version of Fig.~\ref{fig:AFB}, which overlays the $U(1)_{B-L}$ and BLR models,   
for a luminosity of 3000 fb$^{-1}$ corresponding to the final result for the High-Luminosity LHC
(HL-LHC) run \cite{Gianotti:2002xx}. The purple region is the overlap of errors between the two models. One can see that there are areas where the errors do not overlap and, by looking at the entire invariant mass distribution, a detailed statistical analysis may in principle differentiate between these two models at this luminosity,
although we leave this task to the experimental collaborations.

\subsection{Higgs Final States}
\begin{figure}
	\includegraphics[width=0.7\linewidth]{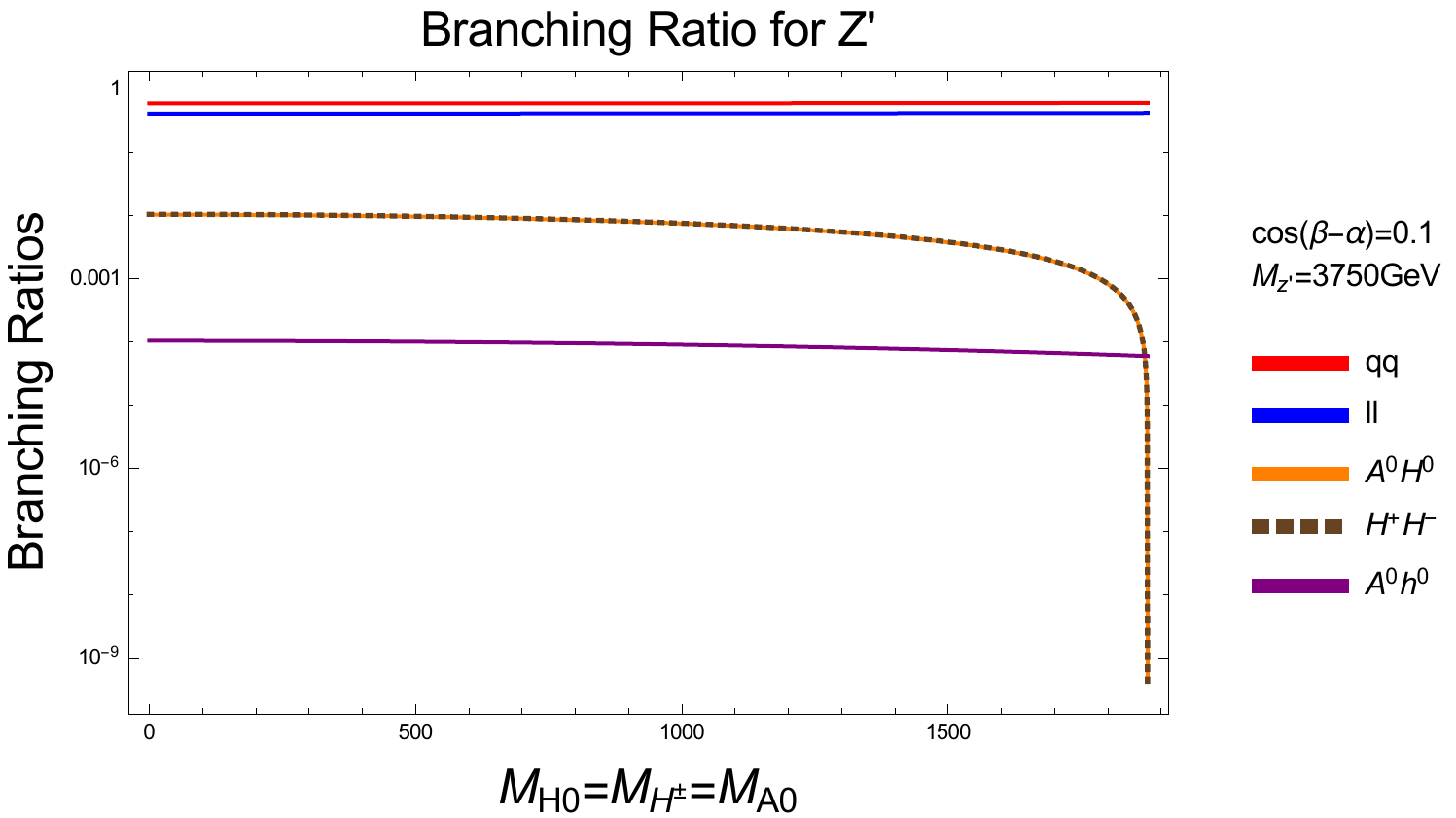} 
	\caption{\footnotesize BRs of a $Z'$ in the BLR model as a function of degenerate $A_0$, $H^0$ and $H^\pm$ masses. Here, $M_{Z'}=3750$ GeV and $\cos(\beta-\alpha)=0.1$.}
	\label{fig:BR}
\end{figure}
An alternative way of singling out the BLR nature of a $Z'$ signal established via DY studies would the one pursuing the isolation of its exotic decays, i.e., into non-SM objects.
Under the enforced assumption of heavy neutrinos, additional CP-odd Higgs boson and all sparticles being (much) heavier than the $Z'$, the latter would include those into all possible MSSM-like (pseudo)scalar states pertaining to the Higgs sector of the BLR model, which, as discussed while commenting Tab.~\ref{particlecontent}, is notably different from those of the   $U(1)_{B-L}$ scenario. In particular, in presence of CP conservation, the following decay channels would be allowed in the BLR framework: $Z'\to A^0h^0$, $A^0H^0$ and $H^+H^-$.
These are presented for the usual $Z'$ benchmark, assuming $\cos(\beta-\alpha)=0.1$ (so as to comply with LHC data from Higgs studies), in Fig.~\ref{fig:BR}, for representative values of the Higgs boson masses. While the corresponding BRs are always subleading (of ${\cal O}(10^{-4})$ to ${\cal O}(10^{-2})$) with respect to those of the decays into SM fermions, the (on-shell) $Z'$ cross section is 2.46 fb, so that HL-HLC luminosities should render the extraction of these decay modes possible, whichever the final decay patterns of the Higgs bosons involved.

\section{Conclusion}
\label{conclusion}
$SO(10)$ GUTs have the remarkable property that they predict right-handed neutrinos,
making neutrino mass inevitable. $SO(10)$ is also a rank 5 gauge group, which implies that any rank preserving
GUT breaking sector will lead to an extra Abelian factor in the low energy effective theory,
which protects right-handed neutrinos from gaining mass. 
If the rank is broken at the TeV scale, then there will be a $Z'$ and massive right-handed neutrinos
possibly observable at the LHC. 

We have considered $SO(10)$ motivated $Z'$ models.
In particular we have focussed on the breaking pattern in Eq.~\ref{BLR00},
where the final breaking scale in Eq.~\ref{BLR}, of the \URxUBL\ Abelian subgroup into the hypercharge $U(1)_Y$ of
the SM, may be around
the TeV scale without spoiling gauge unification, within the accuracy of our one-loop analysis.
The SUSY version of the  \URxUBL\  (BLR) model permits a
linear  seesaw mechanism for neutrino mass generation.

After defining the BLR model particle content and giving the relevant $Z'_{BLR}$ and Higgs couplings, 
we have focussed on 
the discovery prospects of the $Z'_{BLR}$ at the LHC,
its decay into Higgs states, and the forward-backward asymmetry as a diagnostic for discriminating it from 
the $Z'_{BL}$ of the $U(1)_Y\times U(1)_{B-L}$ model. It is noteworthy that the $Z'_{BL}$ of the $B-L$ model has purely vector couplings to 
quarks and leptons, making the forward-backward asymmetry a powerful discriminator, as we have discussed.
In general, we have set out to test whether such models can be
disentangled at past (like LEP/SLC) and present (like LHC) machines, assuming that the SUSY scale is higher than the 
$Z'_{BLR}$ mass.

%We have seen that the BLR model and the $U(1)_{\chi}$ model are identical twins at the GUT scale,
%since the generators $Q_{LR}$ and $T_{\chi}$ in Eqs.~\ref{ZpBLR} and~\ref{Tchi} are identically equal
%if the properly normalised gauge couplings are unified. However at low energy the gauge couplings will diverge and the two 
%twins become non-identical. With this aim in mind, 
%we have studied the RGE evolution at the one-loop level of the BLR model from the GUT scale down to the EW one,
%taking into account gauge kinetic mixing which we have shown to be negligible.
%The small deviation of the gauge couplings $\alpha_R$ and $\alpha_{BL}$ at low energies completely accounts for the 
%phenomenological differences between the $Z'$ in the BLR and the $U(1)_{\chi}$ models.

%We have found that the one-loop RGE analysis of the  \URxUBL\ BLR model
%predicts a value of the hypercharge coupling $\alpha_Y$ which is approximately 
%consistent with the one most  precisely measured (at $M_Z$) by experiment and that differs from the corresponding MSSM prediction by some 4\% (assuming the QCD and weak couplings are the same in both scenarios). 
%Given this, in a future work it would be interesting to consider the RGEs to two-loop accuracy, including threshold effects,
%in order to better distinguish the prospects for gauge coupling unification in the BLR model vs the MSSM,
%and also to more accurately resolve the  $Z'$ in the BLR model and the $U(1)_{\chi}$ models,
%bearing in mind that experimental sensitivities will improve in the years to come.

Having determined the parameters of the BLR model to one-loop accuracy
at the TeV scale, we have examined the feasibility of the LHC to extract a $Z'_{BLR}$ 
signal. We have shown that $Z'_{BLR}$ mass values just below the current sensitivity of the LHC can easily be accessed by the end of Run 2 in standard DY searches exploiting electron and muon final states. Furthermore, we have made a detailed investigation of $A^*_{\rm FB}$ (i.e., the reconstructed forward-backward asymmetry) of these di-lepton final states and shown that it may be possible to distinguish the 
$Z'_{BLR}$ of the  \URxUBL \ from the $Z'_{BL}$ of the 
$U(1)_Y \times U(1)_{B-L}$ case, assuming HL-LHC luminosities.
This is probably the main new result of this paper.
%However to separate  \URxUBL \ from the $U(1)_\chi$ case would seem to require an $e^+e^-$ machine. This phenomenology is dictated by the fact that the $Z'$ of the  $U(1)_Y \times U(1)_{B-L}$ scenario is purely vector, unlike the \URxUBL\  and $U(1)_\chi$ ones, which have very similar vector and axial couplings.

We have also considered the $Z'_{BLR}$ decays into MSSM-like Higgs bosons, which would include $Z'_{BLR}\to A^0h^0$, $A^0H^0$ and $H^+H^-$, but excluding possible decays into $\chi_R^1$ and $\chi_R^2$ bosons which we assume to be too heavy to be produced.
While the Higgs decay rates are always small, from percent to fraction of permille level, compared to those into SM leptons and quarks,  HL-HLC luminosities should render the extraction of all of these signals feasible. 
Though such decays are often neglected in the literature, they provide an additional Higgs production mechanism,
possibly the dominant mechanism on the $Z'_{BLR}$ resonance at an $e^+e^-$ collider, and a crucial test of the gauge structure 
of the model in the 2HDM versions of the models that SUSY demands.

%In conclusion, $SO(10)$ GUTs are well motivated by neutrino mass and allow the possibility of having both right-handed neutrinos and a $Z'$ within the reach of the LHC, where gauge coupling unification suggests SUSY, possibly broken at a higher scale.
%We have seen that the two most favoured $SO(10)$ motivated models, the BLR model based on  \URxUBL \ and the $U(1)_\chi$ model will be 
%very difficult to resolve. To distinguish these two models will require firstly a discovery of a $Z'$, which will be perhaps possible at the HL-LHC, then a detailed study of the forward-backward asymmetry which would require luminosities beyond that envisaged at the HL-LHC, and would most likely require
%an $e^+e^-$ collider. Since 
%the phenomenological differences between the two models arise only from RGE running, the results in this paper motivate a two-loop analysis including threshold effects.

\section*{Acknowledgements}
SM is supported in part through the NExT Institute. All authors acknowledge support from the grant H2020-
MSCA-RISE-2014 n. 645722 (NonMinimalHiggs), the European Union's Horizon 2020 Research and Innovation programme under Marie Sk\l{}odowska-Curie grant agreements Elusives ITN No.\ 674896 and InvisiblesPlus RISE No.\ 690575.
SJDK would like to thank Luigi Delle Rose and Ronald Rodgers for useful discussions.
Finally, we thank Juri Fiaschi for discussion and numerical help.

\appendix

\section{Linear Seesaw}
\label{A}
\noindent
The linear seesaw is similar to an inverse seesaw, but with $\mu \rightarrow 0$ and a new term coupling a left-handed 
(LH) neutrino to the scalar singlet $S$:
\begin{equation}
\begin{pmatrix}
0 & Y v & F v_L \\
Y^T v & 0 & \tilde{F} v_R \\
F^T v_L & \tilde{F} ^T v_R & 0
\end{pmatrix}\equiv
\begin{pmatrix}
0 & m_D & \epsilon \\
m_D ^T & 0 & M_\chi \\
\epsilon ^T & M_\chi ^T & 0
\end{pmatrix}.
\label{eq:App1}
\end{equation}
Each element here corresponds to a $3\times 3$ block. Solving this in block diagonal form, assuming $\epsilon \ll m_D \ll M_\chi$, one finds
\begin{equation}
\begin{pmatrix}
M_\chi + m_D ^T m_D M_\chi ^{-1} & 0 & 0 \\
0 & -(M_\chi + m_D ^T m_D M_\chi ^{-1}) & 0 \\
0 & 0 & -\epsilon \frac{m_D ^T}{M_\chi}
\end{pmatrix}.
\end{equation}
So the light and heavy physical masses are
\begin{align}
M_{\nu_L} &= -\epsilon \frac{m_D ^T}{M_\chi} +\textrm{h.c.} \\
M_{N_1} \sim M_{N_2} &\sim M_\chi + m_D ^T m_D M_\chi ^{-1} + \textrm{h.c.}
\end{align}
Here we have the light neutrinos, $\nu _L$, as observed in oscillation experiments, and $N_{1,2}$ are the heavier neutral fermions. The smallness of $\epsilon$ may allow for a low (TeV) scale $M_\chi$, which is a fundamental feature of all low-scale seesaw mechanisms. Unlike the inverse seesaw, we see that $M_{\nu_l}$ is linear in $m_D$, which is proportional to the Yukawa couplings, hence the name ``linear'' seesaw.

%\section{Gauge-Kinetic Mixing}
%\label{sec:appendix_gauge_kinetic_mixing}

\section{RGEs}
\label{sec:appendix_rge}
\begin{table}[h]
	$
	\left.
	\begin{tabular}{c | c | c | c}
	\hspace{0.4cm}
	\phantom{$b_{R,B-L} ^{BLR} = b_{B-L,R} ^{BLR}$}& \phantom{GUT normalisation} & \phantom{$-1/\sqrt{24}$} & \phantom{$M_\mathrm{SUSY} < Q < M_{\mathrm{\rm GUT}}$} \\ [-\normalbaselineskip]
	Coefficient & GUT normalisation & Value & Scale \vspace{-0.065cm}
	\end{tabular}
	\color{white} \right\} $\phantom{Non-Abelian}
	$
	\left.
	\begin{tabular}{c | c | c | c}
	\phantom{Coefficient} & \phantom{GUT normalisation} & \phantom{Value} & \phantom{Scale} \\ [-\normalbaselineskip] \hline
	\multirow{2}{*}{$b_{R,R} ^{BLR}$} & \multirow{2}{*}{1} & 15/2 & $M_\mathrm{SUSY} < Q < M_{\mathrm{\rm GUT}}$\\
	&&13/3&$M_\mathrm{BLR} < Q < M_\mathrm{SUSY}$\\\hline
	
	\multirow{2}{*}{$b_{(B-L),(B-L)} ^{BLR}$} & \multirow{2}{*}{$3/8$} & 27/4 & $M_\mathrm{SUSY} < Q < M_{\mathrm{\rm GUT}}$\\
	&&17/4&$M_\mathrm{BLR} < Q < M_\mathrm{SUSY}$\\ \hline
	
	\multirow{2}{*}{$b_{R,B-L} ^{BLR} = b_{B-L,R} ^{BLR}$} & \multirow{2}{*}{$\sqrt{3/8}$} & $-\sqrt{3/8}$ & $M_\mathrm{SUSY} < Q < M_{\mathrm{\rm GUT}}$\\
	&&$-1/\sqrt{24}$&$M_\mathrm{BLR} < Q < M_\mathrm{SUSY}$ \vspace{-0.065cm}\\	
	\end{tabular}
	\right\}$ Abelian
	$\left. \hspace{0.75cm}
	\begin{tabular}{c | c | c | c}
	\phantom{$b_{R,B-L} ^{BLR} = b_{B-L,R} ^{BLR}$}& \phantom{GUT normalisation} & \phantom{$-1/\sqrt{24}$} & \phantom{Scale} \\ [-\normalbaselineskip] \hline
	\multirow{3}{*}{$B_3 ^{BLR}$} & 1&-3 & $M_\mathrm{SUSY} < Q < M_{\mathrm{\rm GUT}}$\\
	&1&-7&$M_\mathrm{BLR} < Q < M_\mathrm{SUSY}$\\
	&1&-7&$M_\mathrm{EW} < Q < M_\mathrm{BLR}$\\ \hline
	
	\multirow{3}{*}{$B_2 ^{BLR}$} & 1&1 & $M_\mathrm{SUSY} < Q < M_{\mathrm{\rm GUT}}$\\
	&1&-19/6&$M_\mathrm{BLR} < Q < M_\mathrm{SUSY}$\\
	&1&-19/6&$M_\mathrm{EW} < Q < M_\mathrm{BLR}$\\		
	\end{tabular}
	\right\}
	$Non-Abelian
	\label{tab:Beta_function_coefficients_table}
	\caption{Beta function coefficients for Abelian and non-Abelian gauge groups in the BLR model}
	\end{table}

	\noindent
	Beta functions for the non-Abelian and Abelian groups, respectively, are \cite{Bertolini:2009qj}
	\begin{equation}
	\frac{\textrm{d} g_a}{\textrm{d} t} = \frac{B_a g_a^3}{16 \pi ^2}~,~~\frac{d g_{lm}}{dt} = \frac{g_{lk}}{16 \pi^2} b_{ij}g_{ik} g_{jm},
	\label{eq:beta_functions}
	\end{equation}
	where the index $a$ runs over the non-Abelian groups $SU(2)_L$ and $SU(3)_c$, $a=2,3$ and $(i,j,k,l,m)$ run over the $U(1)_R ,~U(1)_{B-L}$, and mixed $U(1)_R \times U(1)_{B-L}$ and $U(1)_{B-L} \times U(1)_R$ groups, $(i,j,k,l,m)=(R,B-L)$ and Einstein summation convention is assumed. For our RGE section, we make a rotation on the coupling matrix $G$, such that it is set in upper triangular form \cite{Coriano:2015sea}
	\begin{equation}
	G = \begin{pmatrix}
	g_{11} & g_{12} \\
	g_{21} & g_{22}
	\end{pmatrix}
	\end{equation}
	\begin{equation}
	\tilde{G} = G O^T _R =
	\begin{pmatrix}
	g & \tilde{g} \\
	0 & g'
	\end{pmatrix}
	=\left(
	\begin{array}{cc}
	\frac{g_{11} g_{22}-g_{12} g_{21}}{\sqrt{g_{21}^2+g_{22}^2}} & \frac{g_{11} g_{21}+g_{12} g_{22}}{\sqrt{g_{21}^2+g_{22}^2}} \\
	0 & \sqrt{g_{21}^2+g_{22}^2} \\
	\end{array}
	\right)
	\end{equation}
One may consequently find the RGE in terms of $g,~g',~\tilde{g}$ by differentiating these expressions and then replacing the differentials $d g_{ij} /dt$ with the beta functions as calculated with eq. \ref{eq:beta_functions}, then replacing $g_{11},g_{12},g_{22}$ in terms of $g,~g',~\tilde{g}$.

\end{document}